\documentclass[preprintnumbers,showkeys]{revtex4}
\usepackage{amsmath,amsfonts,amssymb,amscd,amsxtra,amsthm}
\usepackage{graphicx}
\usepackage{epstopdf}
\usepackage{bm}
\usepackage{orcidlink}
\usepackage{stackengine}
\usepackage{braket}
\usepackage{ragged2e}

\begin{document}
\preprint{PKNU-NuHaTh-2026}
\title{Lesser Green's Function and Chirality-Reduced Entropy\\ via the In-Medium NJL Model}

\author{Seung-il Nam$^*$\,\orcidlink{0000-0001-9603-9775}}
\email[Email: ]{sinam@pknu.ac.kr}
\affiliation{Department of Physics, Pukyong National University (PKNU), Busan 48513, Republic of Korea}
\affiliation{Asia Pacific Center for Theoretical Physics (APCTP), Pohang 37673, Republic of Korea}
\date{\today}
\begin{abstract}
We investigate chiral symmetry restoration in hot and dense quark matter using a correlator-based chirality-reduced entropy in the in-medium Nambu--Jona-Lasinio (NJL) model. Starting from the lesser Green's function $G^{<}(k)$ in the real-time formalism, we construct the equal-time correlation matrix $C(\bm{k})$ and define the left-handed reduced correlator $C_L(\bm{k}) = P_L C(\bm{k}) P_L$. The corresponding von Neumann entropy, $S_\chi = -\mathrm{Tr}\left[C_L \ln C_L + (1-C_L)\ln(1-C_L)\right]$, characterizes the mixedness of the chirality-reduced subsystem. We show that the reduced correlator retains a nontrivial helicity structure and must therefore be described by its full eigenvalue spectrum rather than by a single scalar occupation probability. The self-consistent dynamical quark mass $M_q(T,\mu_q)$ reproduces the expected QCD-like phase structure, with a second-order transition in the chiral limit and a smooth crossover for finite current quark mass. The chirality-reduced entropy correlates with chiral restoration but is not itself an order parameter; instead, it provides complementary information via the full spectrum of the reduced correlator. Our numerical results show that $S_\chi$ exhibits characteristic nontrivial behavior across the chiral transition region and serves as an information-theoretic diagnostic of reduced chiral-sector mixedness.
\end{abstract}
\maketitle
\section{Introduction}
The spontaneous breakdown (SBCS) and restoration of chiral symmetry are central aspects of Quantum Chromodynamics (QCD), playing key roles in hadron mass generation and in the formation of the quark--gluon plasma (QGP) at high temperature or density~\cite{Nambu:1961fr,Klevansky:1992qe,Hatsuda:1994pi,Buballa:2003qv,Fukushima:2010bq}. In the QCD vacuum, chiral symmetry is dynamically broken, leading to a large constituent quark mass and the emergence of light pseudoscalar mesons as Nambu--Goldstone bosons. As temperature or baryon density increases, the quark condensate $\langle\bar{q}q\rangle$ melts, the constituent mass decreases, and chiral symmetry is partially restored. Lattice QCD simulations indicate that this restoration occurs through a smooth crossover at vanishing chemical potential for physical quark masses~\cite{Aoki:2006we,Borsanyi:2010cj}. Effective approaches such as the Nambu--Jona-Lasinio (NJL), Polyakov-NJL (PNJL), and instanton-based models reproduce the qualitative features of the QCD phase diagram~\cite{Klevansky:1992qe,Hatsuda:1994pi,Ratti:2005jh,Fukushima:2008wg,Nam:2009nn}.

Recent developments have applied concepts from quantum information theory to strongly interacting systems, using entropy-based observables to characterize correlations and mixedness~\cite{Vedral:2002zz,Amico:2007ag, Calabrese:2009qy,Casini:2014aia}. In QCD and related field theories, such ideas have been explored in connection with confinement, deconfinement, topology, spin structures, and high-energy observables, including entanglement entropy in deep-inelastic scattering, entropy production in hadronic final states, and QCD evolution of entanglement-related quantities~\cite{Levin:2006zz,Ryu:2006bv, Kharzeev:2017qzs,Zhang:2021hra,Hentschinski:2024gaa,Nam:2025hei}.
These developments show that information-theoretic quantities can provide useful probes of correlations that are not always directly visible in conventional local order parameters.

The present work is positioned differently from studies based on spatial bipartite entropy, small-$x$ partonic entropy, or color-sector entropy. We focus instead on an internal chiral reduction of the fermionic equal-time correlator. Starting from the real-time lesser Green's function, we construct the correlator $C(k)$, project it onto the left-handed subspace as $C_L(k)=P_L C(k)P_L$, and define the chirality-reduced entropy from the eigenvalue spectrum of $C_L(k)$. Thus, $S_\chi$ is designed to quantify the mixedness of a chirality-reduced Gaussian subsystem. It is correlated with chiral restoration, but it is not an order parameter and should not be identified with the chiral condensate or the dynamical quark mass. In this sense, the present correlator-based entropy provides a complementary perspective to conventional symmetry-based observables.

In this work, we investigate chiral symmetry restoration in hot and dense quark matter using a correlator-based chirality-reduced entropy within the in-medium NJL model. Starting from the lesser Green's function $G^{<}(k)$ in the real-time formalism~\cite{KadanoffBaym1962,Keldysh:1964ud}, we construct the equal-time correlation matrix
\begin{equation*}
C(\bm{k}) = i\int \frac{dk^0}{2\pi} G^{<}(k),
\end{equation*}
and define the left-handed reduced correlator $C_L(\bm{k}) = P_L C(\bm{k}) P_L$. The corresponding von Neumann entropy,
\begin{equation*}
S_\chi=-\mathrm{Tr}\left[C_L\ln C_L+(1-C_L)\ln(1-C_L)\right],
\end{equation*}
is used to quantify the mixedness of the chirality-reduced subsystem.

Within the mean-field NJL framework, the reduced correlator retains a nontrivial helicity structure and must therefore be described by its full eigenvalue spectrum rather than by a single scalar occupation probability. The self-consistent dynamical quark mass $M_q(T,\mu_q)$ reproduces the expected QCD-like phase structure, with a second-order transition in the chiral limit and a smooth crossover for finite current quark mass. The resulting entropy is correlated with chiral restoration but is not an order parameter; instead, it reflects the structure of the reduced correlator beyond local one-point observables.

Conceptually, this quantity differs from the conventional chiral condensate $\langle\bar q q\rangle$. While the condensate is a local expectation value characterizing the magnitude of symmetry breaking, $S_\chi$ is defined through the spectrum of a reduced fermionic correlation matrix restricted to a chiral subspace. It therefore characterizes the mixedness of the chiral sector rather than the strength of symmetry breaking itself.

Most existing entropy-related studies in QCD focus on spatial bipartite entropy, R\'enyi entropies, or color-sector observables in Euclidean or lattice formulations. In contrast, the present work considers an entropy associated with an internal chiral degree of freedom, constructed from the real-time lesser Green's function via the corresponding equal-time correlator. This provides a complementary probe of chiral dynamics that is not directly captured by conventional symmetry-breaking observables.

This paper is organized as follows. In Sec.~II, we outline the finite-temperature NJL framework and the real-time formalism for the lesser Green's function. In Sec.~III, we construct the equal-time correlator and derive the chirality-reduced correlator. Section~IV presents the reduced correlator in the chiral basis and its eigenvalue structure. In Sec.~V, we introduce the von Neumann chirality-reduced entropy and discuss its interpretation. Numerical results are presented in Sec.~VI, followed by conclusions in Sec.~VII.

\section{In-medium NJL Model}
The Nambu--Jona-Lasinio (NJL) model provides an effective description of dynamical chiral symmetry breaking based on a chirally symmetric four-fermion interaction. Its Lagrangian is given by
\begin{equation}
\mathcal{L}_{\text{NJL}} = \bar{q}(i\gamma^{\mu}\partial_{\mu} - m_q)q 
+ G\left[(\bar{q}q)^2 + (\bar{q}i\gamma_5\bm{\tau}q)^2\right].
\label{eq:NJL}
\end{equation}
Here, $q=(u,d)^T$ denotes the light quark fields, $m_q$ is the current quark mass, and $G$ is the coupling constant. The interaction term preserves the global $SU(2)_L \times SU(2)_R$ symmetry in the chiral limit.

In the present work, we employ the minimal two-flavor NJL model with scalar and pseudoscalar interactions. This choice isolates the role of the dynamically generated constituent mass in the chirality-reduced correlator and the associated entropy. Additional interaction channels, such as vector interactions, primarily modify the medium dependence and are not essential for the mechanism under consideration.

Dynamical mass generation arises through the formation of a quark condensate, leading to the constituent quark mass
\begin{equation}
M_q = m_q - 2G \langle \bar{q}q \rangle,
\label{eq:MQQ}
\end{equation}
which is obtained self-consistently from the finite-temperature gap equation,
\begin{align}
M_q(T,\mu_q) = m_q - 2G\langle\bar{q}q\rangle, \quad
\langle\bar{q}q\rangle = -4N_cN_f\int^\Lambda_0\frac{d^3\vec{k}}{(2\pi)^3}\frac{M_q}{E_k}
\big[1-n_+-n_-\big],
\label{eq:gap_Tmu}
\end{align}
where $E_k^2 = |\vec{k}|^2 + M_q^2$, and the Fermi--Dirac distribution functions are defined as
\begin{equation}
n_F(E_k\pm\mu_q)=\frac{1}{e^{(E_k\pm\mu_q)/T}+1}\equiv n_\mp .
\end{equation}
The NJL parameter set used in this work is listed in Table~\ref{tab:NJL_params}.

\setlength{\tabcolsep}{15pt}
\begin{table}[b]
\centering
\begin{tabular}{ccc}
\hline\hline
$m_q$ [MeV] & $\Lambda$ [MeV] & $G\Lambda^2$ \\
\hline
5.25 & 631.4 & 2.14 \\
\hline\hline
\end{tabular}
\caption{NJL parameter set used in the present study.}
\label{tab:NJL_params}
\end{table}

\begin{figure}[t]
\topinset{(a)}{\includegraphics[width=8cm]{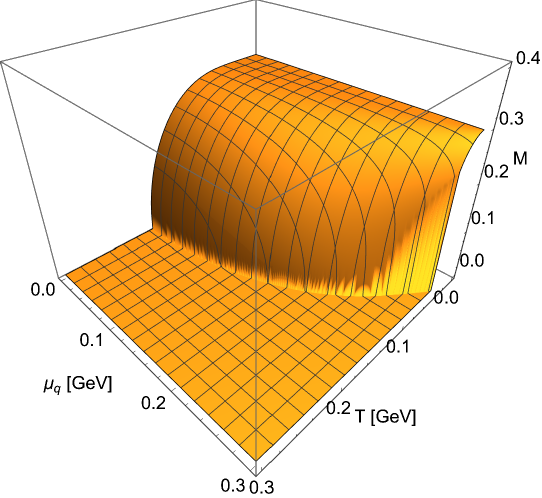}}{-0.4cm}{0.2cm}
\topinset{(b)}{\includegraphics[width=8cm]{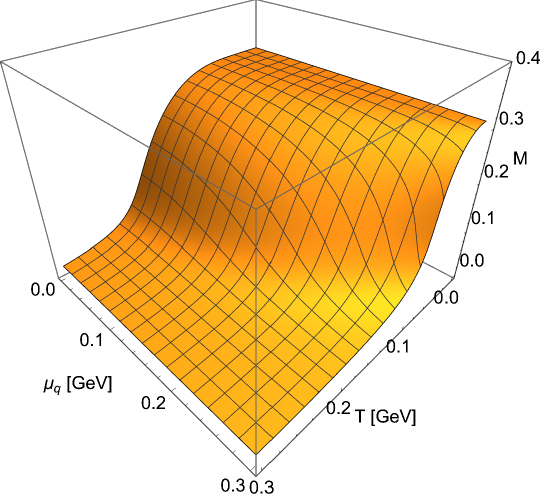}}{-0.4cm}{0.2cm}
\hspace{0.5cm}
\caption{Dynamical quark mass $M_q$ as a function of temperature $T$ and chemical potential $\mu_q$ for $m_q=0$ (a) and $m_q=5.25$ MeV (b).}
\label{FIG1}
\end{figure}

Fig.~\ref{FIG1} shows the temperature and chemical-potential dependence of the dynamical quark mass $M_q(T,\mu_q)$ obtained from Eq.~(\ref{eq:gap_Tmu}). In the chiral limit ($m_q=0$), the constituent mass decreases continuously with temperature at $\mu_q=0$, signaling a second-order transition, while at higher density the transition becomes first order with a critical end point separating the two regimes.

For finite current quark mass ($m_q=5.25~\mathrm{MeV}$), chiral symmetry is explicitly broken, and the phase transition is replaced by a smooth crossover. In this case, $M_q$ decreases gradually with increasing temperature and density, remaining finite even at high temperatures. This behavior is consistent with lattice QCD results and confirms that the NJL model captures the qualitative features of chiral symmetry restoration.

A variation of the model parameters within reasonable ranges, $\Lambda = (630 \pm 30)\,\mathrm{MeV}$ and $G\Lambda^2 = 2.1 \pm 0.1$, was found to modify only the overall magnitude of the results without affecting their qualitative behavior or the location of the pseudo-critical region. This confirms the robustness of the present analysis.

\section{Lesser Green's Function and Chiral Correlator Framework}
In the Schwinger--Keldysh real-time formalism~\cite{Keldysh:1964ud}, the greater and lesser Green's functions are defined as
\begin{align}
G^{>}(x,y) = -i\langle q(x)\bar{q}(y)\rangle, \quad
G^{<}(x,y) = i\langle \bar{q}(y)q(x)\rangle.
\label{eq:GGGG}
\end{align}
In equilibrium, they are related to the spectral function
$A(p)=i\left[G^R(p)-G^A(p)\right]$
and the Fermi distribution $n_F(p^0)$ through
\begin{align}
G^{<}(p) = i\,n_F(p^0)\,A(p), \quad
G^{>}(p) = -i\,[1-n_F(p^0)]\,A(p).
\label{eq:eq_relation_main}
\end{align}
Thus, $G^{<}(p)$ encodes the occupied part of the spectral density. For the Gaussian entropy construction, however, the relevant quantity is not $G^{<}(k)$ itself but the equal-time correlation matrix
\begin{equation}
C(\bm{k}) \equiv i\int\frac{dk_0}{2\pi}\,G^{<}(k),
\label{eq:C_equal_time_main}
\end{equation}
whose eigenvalues lie in the interval $[0,1]$.

Within the mean-field NJL model, the lesser propagator takes the form
\begin{align}
G^{<}(k) = 2\pi i\Big[
n_+\,\Lambda_+(\bm{k})\,\delta(k_0-E_k)
+ (1-n_-)\,\Lambda_-(\bm{k})\,\delta(k_0+E_k)
\Big],
\label{eq:Glesser_full_final}
\end{align}
where $E_k=\sqrt{\bm{k}^2+M_q^2}$ and $n_\pm=n_F(E_k\mp\mu_q)$. The positive- and negative-energy projectors are
\begin{equation}
\Lambda_\pm(\bm{k}) = \frac{1}{2}\left[1 \pm \frac{\gamma^0(\bm{\gamma}\cdot\bm{k}+M_q)}{E_k}\right],
\label{eq:Lambda_pm_final}
\end{equation}
with
\begin{equation}
\Lambda_\pm^2=\Lambda_\pm,\quad
\Lambda_+\Lambda_-=0,\quad
\Lambda_++\Lambda_-=1.
\end{equation}
Integrating Eq.~(\ref{eq:Glesser_full_final}) over $k_0$ gives the equal-time correlator
\begin{equation}
C(\bm{k})=
n_+\,\Lambda_+(\bm{k})+[1-n_-]\,\Lambda_-(\bm{k}).
\label{eq:Ck_final}
\end{equation}
This correlator provides the starting point for the chirality-reduced construction developed below.

In the chiral limit $M_q\to0$, the Dirac Hamiltonian becomes block diagonal in chirality, so that the left- and right-handed sectors decouple dynamically. For finite $M_q$, the Dirac mass term mixes the two chiral sectors through the off-diagonal structure of the Dirac operator in the Weyl basis. Accordingly, $C(\bm{k})$ contains both thermal occupation information and mass-induced left-right mixing. Detailed derivations are given in Appendices A and B.

\section{Chirality Decomposition and Reduced Correlator}
The Dirac field can be decomposed into left- and right-handed components as
\begin{equation}
q=q_L+q_R, \quad q_{L,R}=P_{L,R}q, \quad P_{L,R}=\frac{1\mp\gamma^5}{2}.
\label{eq:PLPR}
\end{equation}
In the chiral limit $M_q\to0$, the two sectors decouple dynamically, while a finite constituent mass induces left-right mixing.

To quantify the reduced left-handed sector, we define the chirality-projected equal-time correlator
\begin{equation}
C_L(\bm{k}) = P_L C(\bm{k}) P_L
= n_+ P_L \Lambda_+(\bm{k}) P_L + (1-n_-) P_L \Lambda_-(\bm{k}) P_L.
\label{eq:CL_def}
\end{equation}

Working in the Weyl basis, the projectors take the block form
\begin{equation}
\Lambda_\pm(\bm{k})=
\frac12
\begin{pmatrix}
I \mp \dfrac{\bm{\sigma}\cdot\bm{k}}{E_k} & \pm \dfrac{M_q}{E_k} \\
\pm \dfrac{M_q}{E_k} & I \pm \dfrac{\bm{\sigma}\cdot\bm{k}}{E_k}
\end{pmatrix}.
\label{eq:Lambda_block_main}
\end{equation}

Projecting onto the left-handed subspace yields
\begin{align}
P_L \Lambda_+ P_L = \frac12\left(I - \frac{\bm{\sigma}\cdot\bm{k}}{E_k}\right)P_L, \quad
P_L \Lambda_- P_L = \frac12\left(I + \frac{\bm{\sigma}\cdot\bm{k}}{E_k}\right)P_L.
\label{eq:PLLP_correct_main}
\end{align}

Substituting into Eq.~(\ref{eq:CL_def}), we obtain
\begin{equation}
C_L(\bm{k}) = a_k I + b_k\,\bm{\sigma}\cdot\hat{\bm{k}},
\label{eq:CL_general_main}
\end{equation}
with
\begin{equation}
a_k=\frac12(1+n_+-n_-), \quad
b_k=\frac{|\bm{k}|}{2E_k}(1-n_- - n_+).
\label{eq:abk_main}
\end{equation}

The eigenvalues are therefore
\begin{equation}
\lambda_\pm(\bm{k}) = a_k \pm b_k, \quad
\Delta\lambda(\bm{k}) \equiv \lambda_+(\bm{k})-\lambda_-(\bm{k}) = 2b_k.
\label{eq:lambdas_main}
\end{equation}
which satisfy $0 \le \lambda_\pm \le 1$. Hence, the reduced correlator is characterized by its full eigenvalue spectrum rather than by a single scalar occupation probability. In particular, for finite $M_q$, the vacuum limit does not generically yield $\lambda_\pm=0,1$ mode by mode.

\begin{figure}[t]
\topinset{(a)}{\includegraphics[width=8cm]{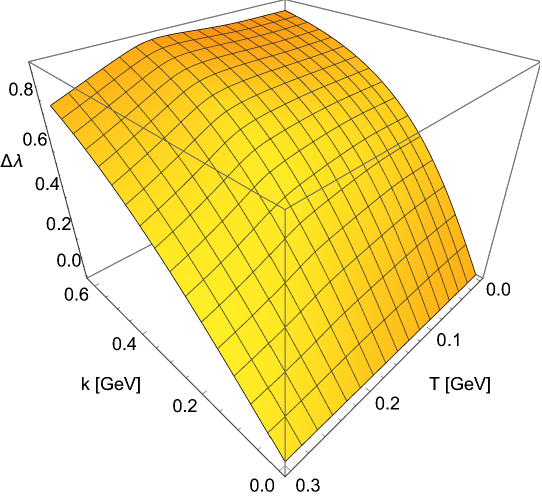}}{-0.4cm}{0.2cm}
\topinset{(b)}{\includegraphics[width=8cm]{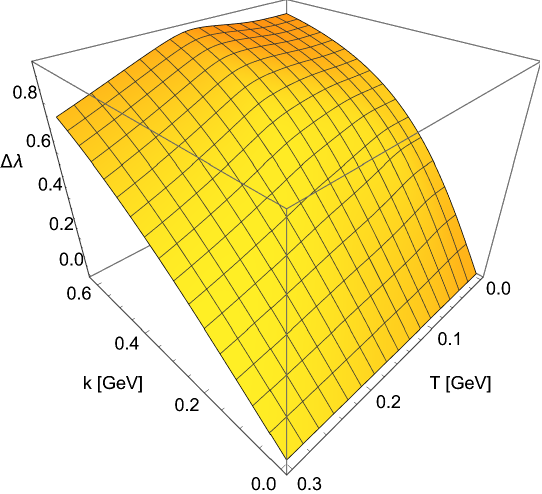}}{-0.4cm}{0.2cm}
\caption{(Color online) Eigenvalue splitting of the chirality-reduced correlator, $\Delta\lambda(\bm{k})=\lambda_+(\bm{k})-\lambda_-(\bm{k})$, as a function of temperature $T$ and momentum $|\bm{k}|$ for $\mu=0$ (a) and $200$ MeV (b) at $m_q\ne 0$.}
\label{FIG4}
\end{figure}

Fig.~\ref{FIG4} shows the eigenvalue splitting $\Delta\lambda(\bm{k})=2b_k$ as a function of temperature and momentum. At low temperature, the splitting remains sizable over a broad momentum range, indicating well-separated eigenmodes. As the temperature increases, thermal broadening drives the system toward $\lambda_+=\lambda_-=\tfrac12$, so that $\Delta\lambda \to 0$, reflecting increased mixedness of the reduced subsystem. Finite density modifies the temperature dependence of the splitting while preserving this overall trend.  

Table~\ref{tab:limits_lambda} presents the limiting forms of the eigenvalues $\lambda_\pm(\bm{k})$ and their splitting $\Delta\lambda$ in several important limits. In the chiral limit $M_q\to0$, one finds $\lambda_\pm(\bm{k})=\{1-n_-,\,n_+\}$ and $\Delta\lambda=1-n_- - n_+$. In the asymptotically high-temperature limit $T\to\infty$, both eigenvalues become $1/2$, implying a vanishing splitting, $\Delta\lambda=0$. In the vacuum limit $(T,\mu_q)\to(0,0)$, the eigenvalues reduce to $\lambda_\pm(\bm{k})=\left(1\pm{|\bm{k}|}/{E_k}\right)/2$, so that $\Delta\lambda={|\bm{k}|}/{E_k}$. Therefore, the table compactly illustrates how the eigenvalue spectrum changes across different physical regimes.

\setlength{\tabcolsep}{15pt}
\begin{table}[b]
\centering
\begin{tabular}{ccc}
\hline\hline
Limit &$\lambda_\pm(\bm{k})$ &$\Delta\lambda$ \\
\hline
$M_q \to 0$&$\{1-n_-,\,n_+\}$&$1-n_- - n_+$\\
$T \to \infty$&$\dfrac{1}{2}$&$0$\\
$(T,\mu_q)\to(0,0)$&$\dfrac{1}{2}\left(1\pm\dfrac{|\bm{k}|}{E_k}\right)$&$\dfrac{|\bm{k}|}{E_k}$\\
\hline\hline
\end{tabular}
\caption{Limiting forms of the eigenvalues $\lambda_\pm(\bm{k})$ and their splitting $\Delta\lambda$.}
\label{tab:limits_lambda}
\end{table}

\section{Von Neumann Entropy and Chirality-Reduced Entropy}
The von Neumann entropy of the left-handed reduced subsystem is defined from the equal-time correlator $C_L$ as
\begin{equation}
S_\chi=-\mathrm{Tr}\left[C_L\ln C_L+(1-C_L)\ln(1-C_L)\right].
\label{eq:Schi_main}
\end{equation}
For a Gaussian fermionic state, this quantity is determined entirely by the eigenvalues of $C_L$. At finite temperature and chemical potential, it measures the mixedness of the left-handed subsystem and generally contains both quantum and thermal contributions. For this reason, it is more appropriate to interpret $S_\chi$ as a chirality-reduced entropy (or chirality-subsystem entropy) rather than as an unqualified entanglement entropy.

Using Eq.~(\ref{eq:Schi_main}), the entropy density becomes
\begin{equation}
\frac{S_\chi}{V}=N_cN_f\int_0^\Lambda \frac{d^3k}{(2\pi)^3}
\sum_{h=\pm}\left[-\lambda_h(\bm{k})\ln\lambda_h(\bm{k})
-(1-\lambda_h(\bm{k}))\ln(1-\lambda_h(\bm{k}))\right],
\label{eq:entropy_main_revised}
\end{equation}
and the corresponding single-mode entropy is
\begin{equation}
s_\chi(\bm{k})=-\sum_{h=\pm}
\left[\lambda_h(\bm{k})\ln\lambda_h(\bm{k})
+(1-\lambda_h(\bm{k}))\ln(1-\lambda_h(\bm{k}))\right].
\label{eq:single_mode_main}
\end{equation}

The entropy vanishes only when the eigenvalues become idempotent, $\lambda_h=0$ or $1$, and it is maximized when $\lambda_+=\lambda_-=\tfrac12$. In the high-temperature limit, $n_\pm\to\tfrac12$, so that $\lambda_\pm\to\tfrac12$ and the entropy reaches its maximal thermal value. In contrast, in the vacuum limit $T\to0$ and $\mu_q=0$,
\begin{equation}
\lambda_\pm(\bm{k})=\frac12\left(1\pm\frac{|\bm{k}|}{E_k}\right),
\label{eq:vacuum_lambdas}
\end{equation}
which is generally non-idempotent for finite $M_q$. Therefore, the chirality-reduced entropy does not necessarily vanish in the zero-temperature vacuum.

We emphasize that $S_\chi$ cannot be expressed as a functional of the chiral condensate $\langle\bar qq\rangle$ alone. The condensate is a local scalar expectation value characterizing the magnitude of symmetry breaking, whereas $S_\chi$ depends on the full spectrum of the reduced correlator $C_L(\bm{k})$. It therefore encodes information on reduced chiral-sector mixedness that is not contained in conventional local order parameters.

Accordingly, the behavior of $S_\chi(T,\mu_q)$ should be interpreted in terms of the temperature- and density-dependent mixedness of the chirality-reduced subsystem. Although it is correlated with chiral restoration, it should not be identified with a pure entanglement measure unless thermal contributions are removed by an appropriate mixed-state entanglement diagnostic.

Before proceeding, we clarify the approximations underlying the present construction. The NJL interaction is treated at the mean-field level, so that the four-fermion operator is replaced by a self-consistent constituent mass obtained from the gap equation. In the real-time formalism, we further employ the narrow-width quasiparticle approximation for the spectral function, such that the lesser propagator takes the on-shell form in Eq.~(\ref{eq:Glesser_full_final}). The entropy is then computed from the equal-time correlation matrix of the corresponding Gaussian fermionic state. Hence, the present $S_\chi$ quantifies the mixedness of the chirality-reduced Gaussian subsystem determined by the two-point correlator. Beyond mean-field, finite spectral widths and non-Gaussian correlations may quantitatively modify entropy and require a more general treatment.

\section{Numerical Results and Discussion}
\begin{figure}[t]
\topinset{(a)}{\includegraphics[width=8cm]{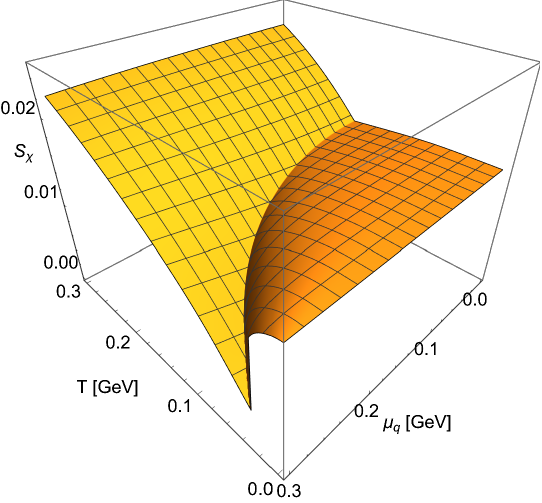}}{-0.4cm}{0.2cm}
\topinset{(b)}{\includegraphics[width=8cm]{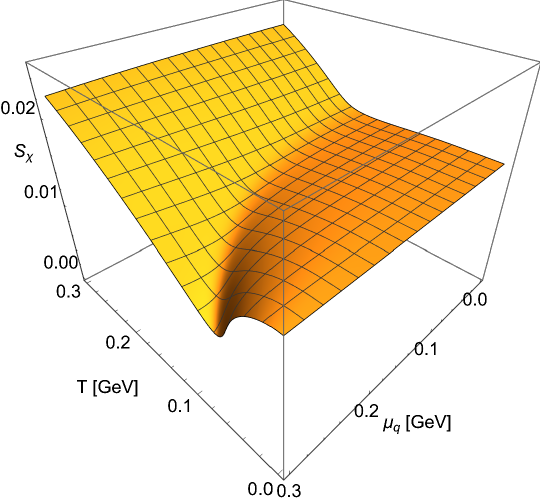}}{-0.4cm}{0.2cm}
\topinset{(c)}{\includegraphics[width=8.5cm]{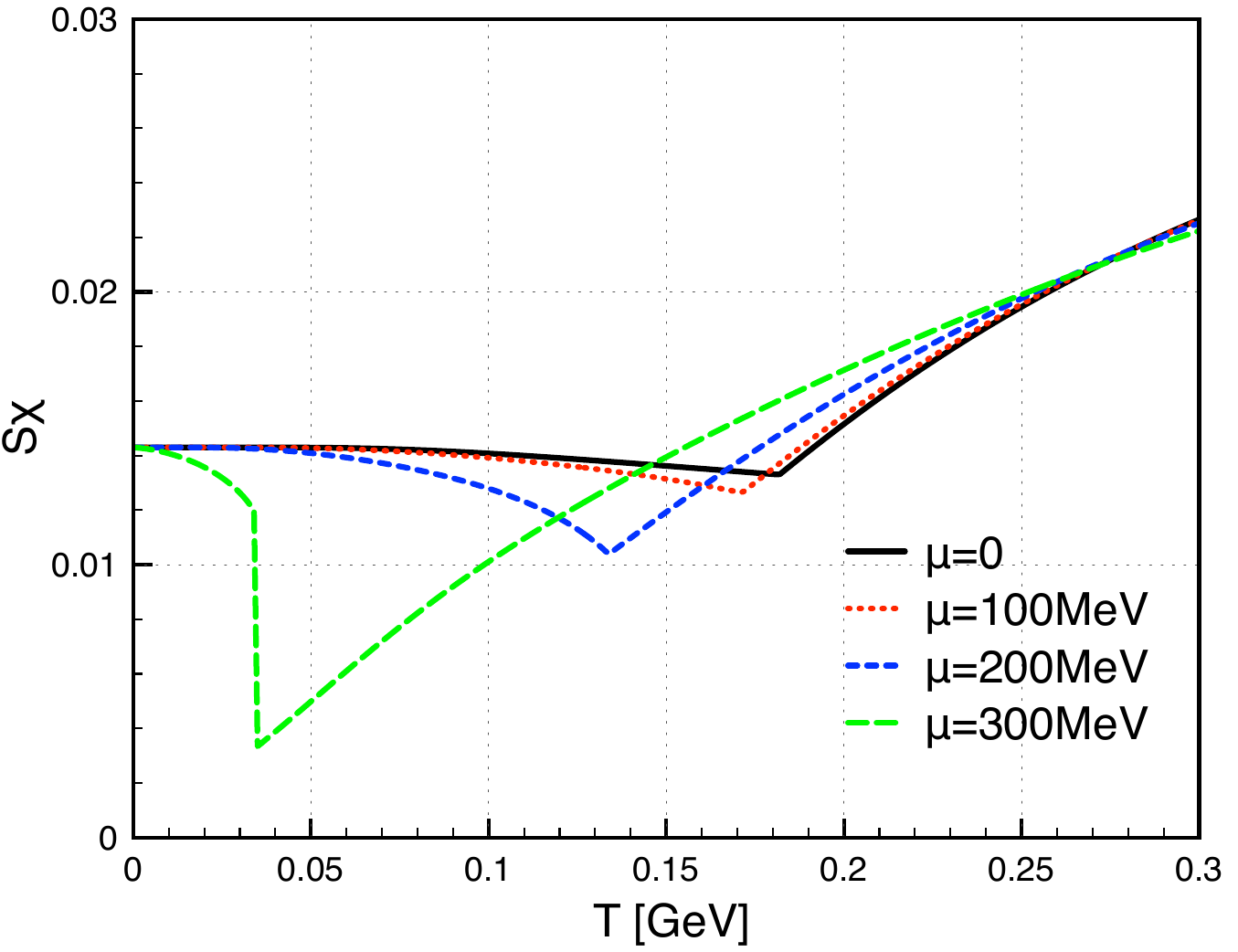}}{-0.4cm}{0.2cm}
\topinset{(d)}{\includegraphics[width=8.5cm]{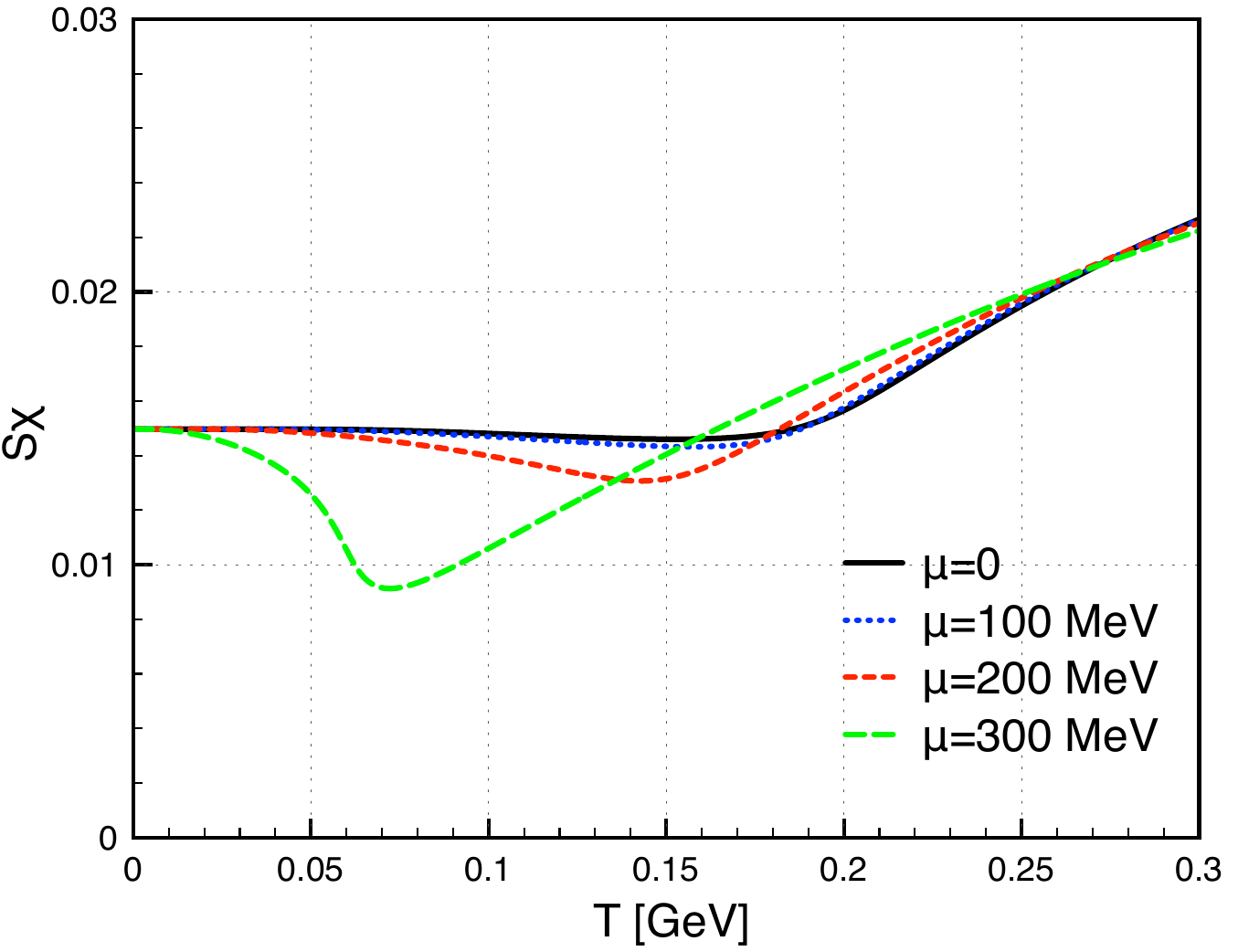}}{-0.4cm}{0.2cm}
\caption{(Color online) Chirality-reduced entropy $S_\chi$ in Eq.~(\ref{eq:entropy_main_revised}) as a function of temperature $T$ and quark chemical potential $\mu_q$ for $m_q=0$ (a) and $m_q=5.25$ MeV (b). Panels (c) and (d) show the corresponding two-dimensional slices at fixed $\mu_q$.}
\label{FIG2}
\end{figure}

We now discuss the numerical behavior of the chirality-reduced entropy $S_\chi$ obtained from Eq.~(\ref{eq:entropy_main_revised}), together with the temperature and density dependence of the dynamical quark mass $M_q(T,\mu_q)$ determined from the NJL gap equation. The finite low-temperature baseline of $S_\chi$ is consistent with the reduced-state interpretation developed in Sec.~V: even in the vacuum limit, the chirality-reduced correlator remains generally non-idempotent for finite $M_q$, so that the reduced entropy does not vanish mode by mode.

Fig.~\ref{FIG2} shows $S_\chi$ as a function of temperature $T$ and quark chemical potential $\mu_q$ for both the chiral limit ($m_q=0$) and the physical current-mass case ($m_q=5.25$ MeV). In both cases, $S_\chi$ exhibits a nontrivial evolution across the chiral transition region. At sufficiently high temperature, the entropy increases as thermal broadening drives the eigenvalues $\lambda_\pm(\bm{k})$ toward $1/2$, thereby enhancing the mixedness of the chirality-reduced subsystem.

A characteristic feature of Fig.~\ref{FIG2} is that, at fixed $\mu_q$, $S_\chi$ does not vary monotonically over the entire temperature range. In particular, for sufficiently large $\mu_q$, the entropy is suppressed in the low-temperature region, followed by a recovery and subsequent increase at higher temperatures. This non-monotonic behavior indicates that the density dependence of $S_\chi$ is intrinsically nontrivial.

\setlength{\tabcolsep}{15pt}
\begin{table}[b]
\centering
\begin{tabular}{c|c|c}
\hline\hline
Parameter & Physical role & Tendency for $S_\chi$ \\
\hline
$\mu_q$ & Fixes fermionic occupation (Fermi-surface formation) & Suppresses \\
$T$ & Induces thermal broadening (occupation smearing) & Enhances \\
$M_q$ & Controls chirality mixing via the Dirac mass & Suppresses \\
\hline\hline
\end{tabular}
\caption{Roles of temperature $T$, chemical potential $\mu_q$, and dynamical mass $M_q$ in determining the chirality-reduced entropy $S_\chi$. The entropy is governed by the competition between Fermi-surface sharpening, thermal broadening, and mass-controlled chirality mixing.}
\label{tab:competition}
\end{table}
The behavior of $S_\chi(T,\mu_q)$ can be understood as the result of a competition among the effects summarized in Table~\ref{tab:competition}. The chemical potential $\mu_q$ determines the fermionic occupation structure through the formation of a Fermi surface, driving a broad momentum region toward nearly idempotent occupation ($n_\pm \simeq 0$ or $1$) and thereby reducing the mixedness of the reduced chiral sector. In contrast, finite temperature induces thermal broadening of the Fermi surface, leading to fractional occupation probabilities and greater mixedness.

The dynamical quark mass $M_q$ provides an additional control through the Dirac structure of the correlator. A large $M_q$ suppresses helicity-dependent mixing and reduces the spread of the eigenvalue spectrum, whereas a decreasing $M_q$ lifts this suppression and allows stronger chirality mixing. The observed behavior of $S_\chi(T,\mu_q)$ is therefore governed by the interplay among Fermi-surface sharpening, thermal broadening, and mass-controlled chirality mixing.

The comparison between the chiral limit and the finite-mass case shows that a finite current quark mass smooths the overall structure of the entropy landscape. In the chiral limit, the variation of $S_\chi$ across the transition region is sharper, whereas for finite $m_q$ the change becomes more gradual, consistent with the crossover nature of the transition in the mass sector.

\begin{figure}[t]
\topinset{(a)}{\includegraphics[width=8cm]{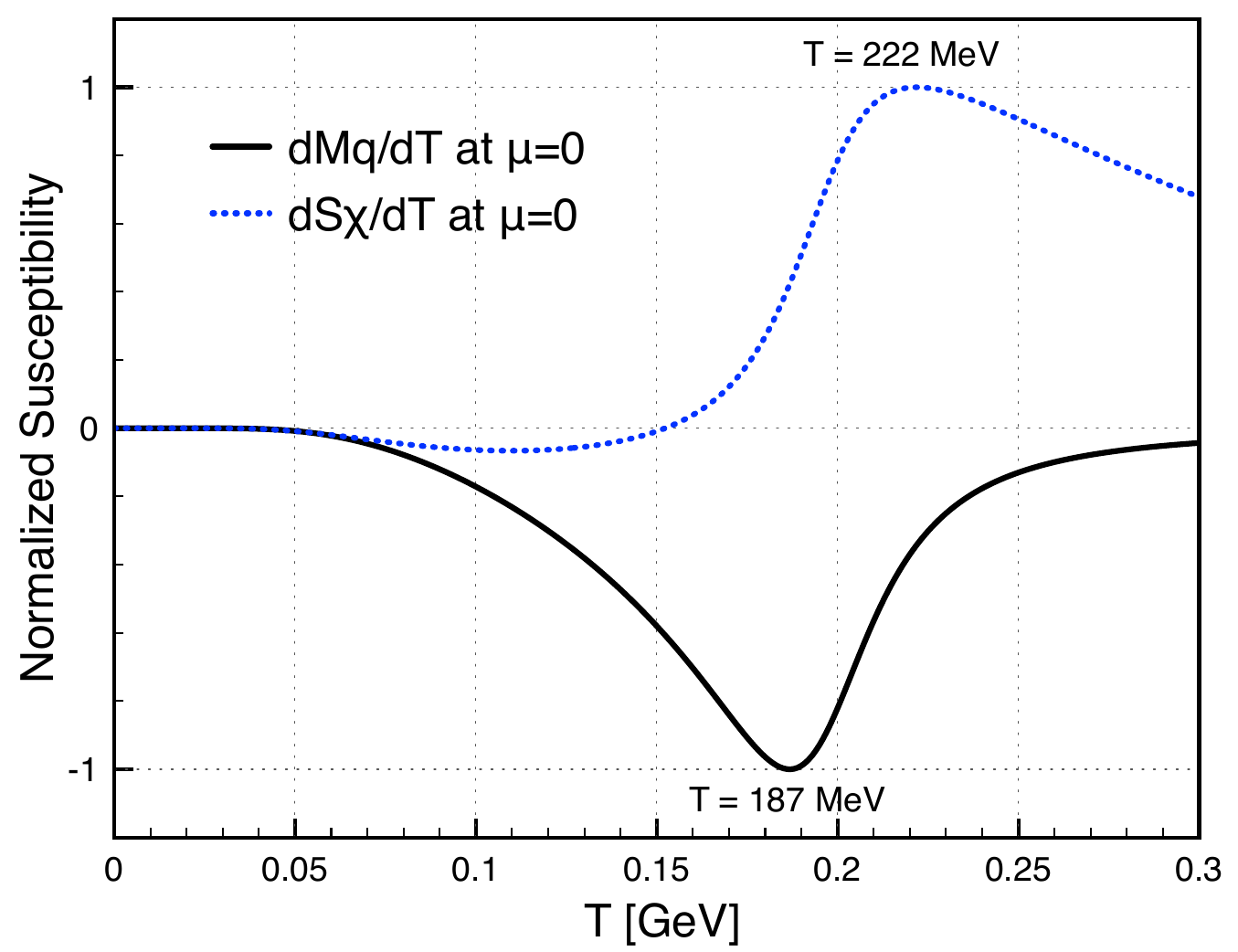}}{-0.4cm}{0.2cm}
\topinset{(b)}{\includegraphics[width=8cm]{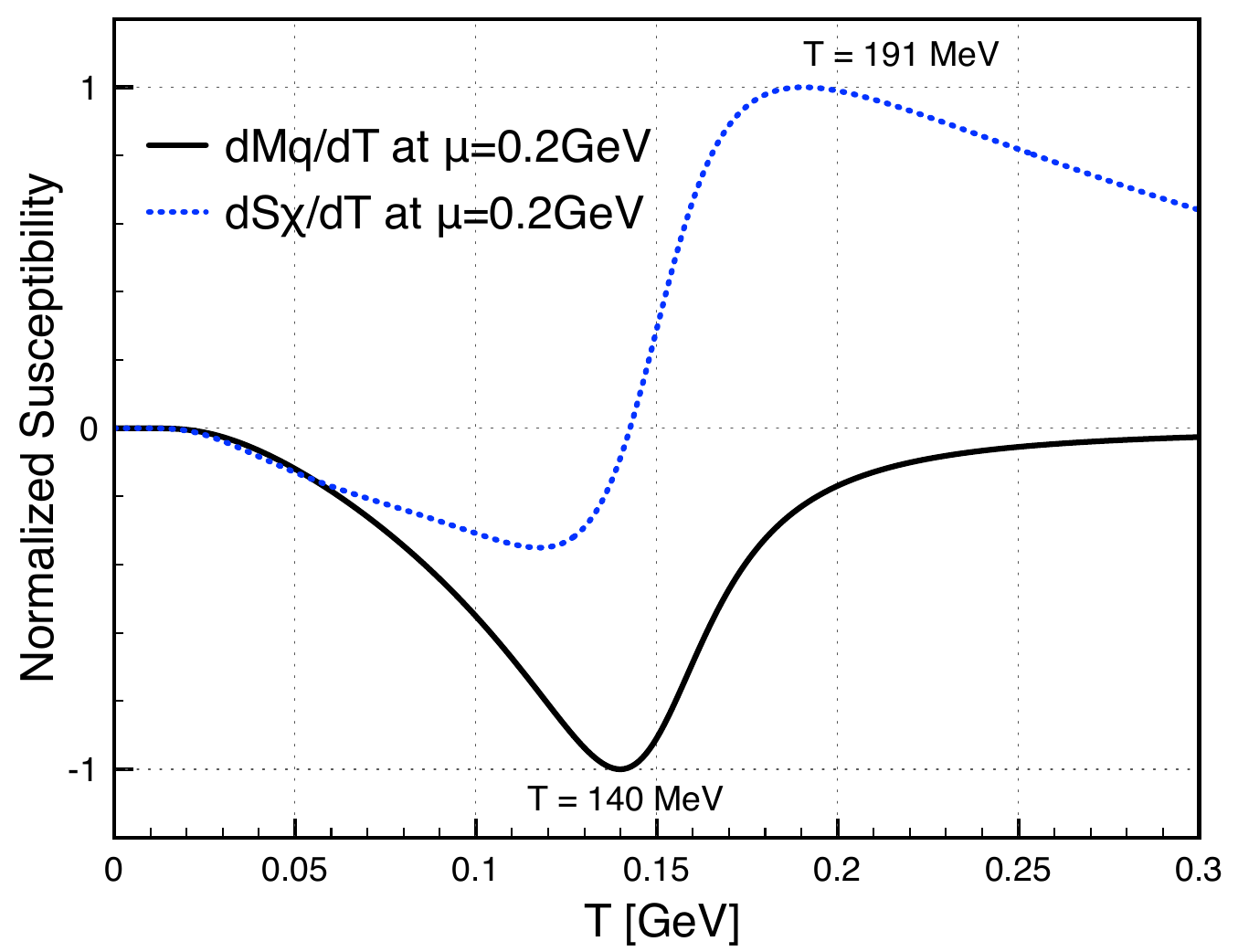}}{-0.4cm}{0.2cm}\\
\topinset{(c)}{\includegraphics[width=8cm]{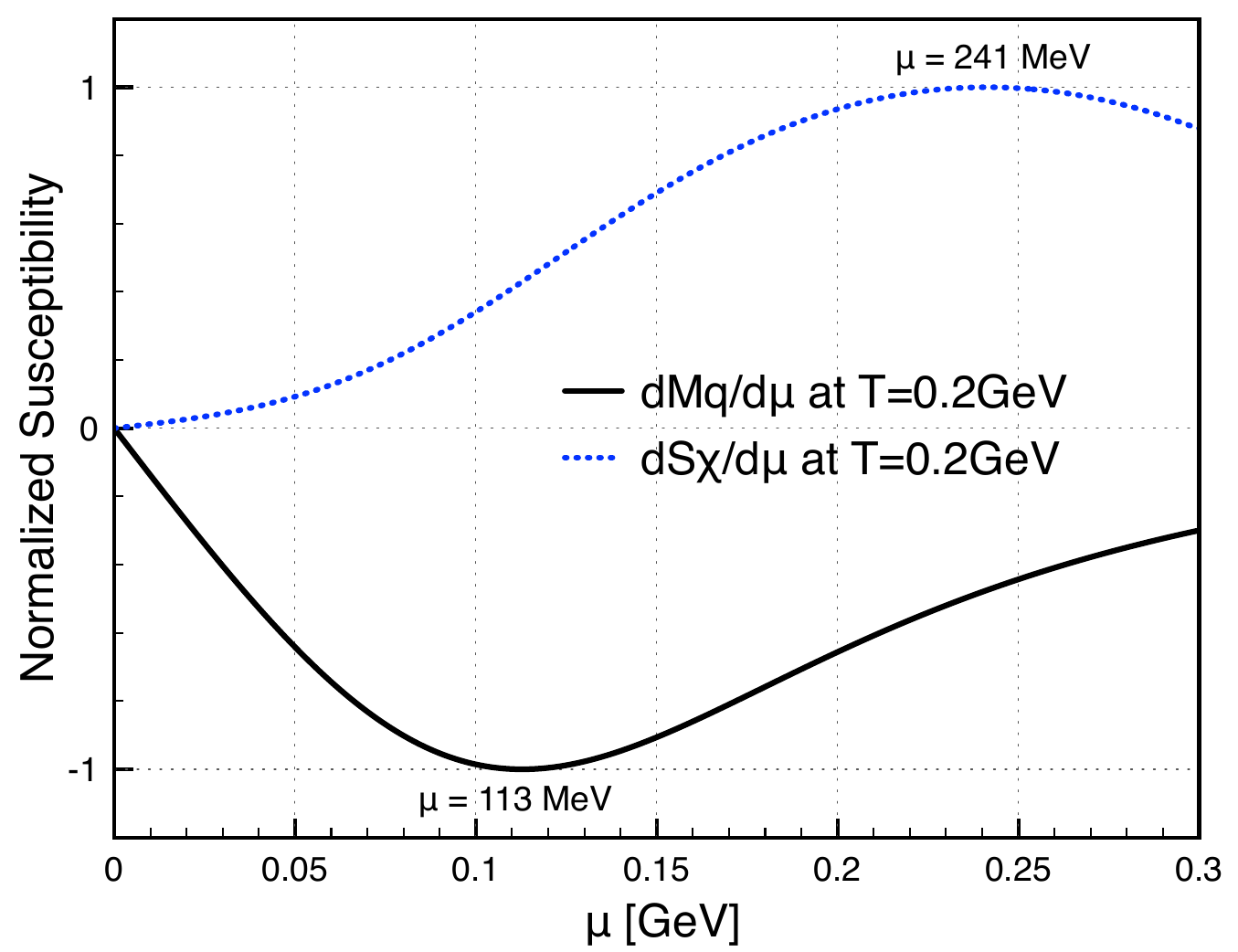}}{-0.4cm}{0.2cm}
\caption{(Color online) Normalized responses of the dynamical quark mass and the chirality-reduced entropy: $dM_q/dT$ and $dS_\chi/dT$ at $\mu_q=0$ (a) and $\mu_q=200$ MeV (b), and $dM_q/d\mu_q$ and $dS_\chi/d\mu_q$ at fixed temperature $T=0.2$ GeV (c). The extrema of the entropy response do not generally coincide with those of the dynamical mass response.}
\label{FIG3}
\end{figure}

To quantify the relation between the entropy and the conventional mass-based signal of chiral restoration, we compare the temperature derivatives of $M_q$ and $S_\chi$ in Fig.~\ref{FIG3}. The extrema of $dM_q/dT$ and $dS_\chi/dT$ do not coincide. At $\mu_q=0$, the minimum of $dM_q/dT$ is located near $T \approx 187$ MeV, while the peak of $dS_\chi/dT$ appears near $T \approx 222$ MeV. At $\mu_q=200$ MeV, the corresponding temperatures shift to approximately $140$ MeV and $191$ MeV, respectively.

This separation has a clear physical origin. The dynamical quark mass $M_q$ probes the onset of symmetry restoration through the melting of the chiral condensate, whereas $S_\chi$ measures the subsequent buildup of mixedness in the fermionic subsystem, which requires sufficient thermal broadening. As a result, the peak of $dS_\chi/dT$ is shifted to higher temperatures compared with that of $dM_q/dT$.

In addition, Fig.~\ref{FIG3}(c) shows that $dM_q/d\mu_q$ is negative, indicating that increasing chemical potential weakens dynamical mass generation, whereas $dS_\chi/d\mu_q$ is positive with a broad maximum, reflecting enhanced sensitivity of the entropy to density. This further demonstrates that $S_\chi$ does not track the mass sector point-for-point, but encodes both mass reduction and density-driven rearrangement of the occupation structure.

Overall, the chirality-reduced entropy provides a correlator-based diagnostic of the transition region in hot and dense quark matter. It is sensitive to both thermal broadening and density-induced occupation restructuring, thereby capturing information beyond conventional symmetry-breaking observables. All results are obtained within the mean-field NJL framework combined with the narrow-width quasiparticle approximation, where the entropy is computed from the equal-time correlation matrix of a Gaussian fermionic state.

\section{Summary and Outlook}
Our results show that chiral symmetry restoration and the mixedness of the chirality-reduced subsystem are closely related, but not identical, even within a mean-field QCD-like model. In this work, we developed a correlator-based framework for analyzing chiral-sector mixedness in hot and dense quark matter within the finite-temperature and density Nambu--Jona-Lasinio (NJL) model. Starting from the lesser Green's function $G^{<}(k)$ in the real-time formalism, we constructed the equal-time correlation matrix $C(\bm{k})$ and defined the left-handed reduced correlator $C_L(\bm{k}) = P_L C(\bm{k}) P_L$. The corresponding von Neumann entropy, $S_\chi = -\mathrm{Tr}\left[C_L \ln C_L + (1-C_L)\ln(1-C_L)\right]$, was used to characterize the mixedness of the chirality-reduced subsystem. In this framework, the reduced correlator retains a nontrivial helicity structure and must therefore be described by its full spectrum of eigenvalues rather than by a single scalar occupation probability.

The self-consistent solutions of the NJL gap equation reproduce the expected QCD-like chiral phase structure, featuring a second-order transition in the chiral limit and a smooth crossover for finite current quark mass. Using these solutions, we evaluated the chirality-reduced entropy density $S_\chi/V$ and found that it exhibits characteristic nontrivial behavior across the chiral transition region. Although $S_\chi$ is correlated with chiral restoration, it is not an order parameter. Instead, it should be interpreted as a correlator-based measure of reduced chiral-sector mixedness, providing complementary information to the conventional chiral condensate via the full spectrum of the reduced correlator.

A central result of the present analysis is that the temperature response of $S_\chi$ does not coincide with that of the dynamical quark mass. In particular, the peak of $dS_\chi/dT$ is shifted relative to the extremum of $dM_q/dT$, indicating that the buildup of reduced-sector mixedness and the rapid melting of the dynamical mass need not occur at the same temperature. At fixed temperature, the chemical-potential responses also show a clear separation between the mass sector and the entropy sector: while $dM_q/d\mu_q$ remains negative, $dS_\chi/d\mu_q$ is positive and develops a broad maximum at intermediate $\mu_q$. These results show that the chirality-reduced entropy probes aspects of chiral dynamics that are not fully captured by conventional symmetry-breaking observables alone.

In this sense, the present study provides an internal chiral-subsystem counterpart to other information-theoretic approaches to QCD. Whereas spatial, partonic, or color-sector entropies probe different partitions of the QCD state, the present $S_\chi$ probes the reduced mixedness associated with chirality projection. This positioning also clarifies the limitation of the observable: $S_\chi$ is not a new order parameter for chiral symmetry restoration, but a correlator-based diagnostic that complements the conventional condensate and mass-based descriptions.

The present framework can be extended in several directions. It would be worthwhile to investigate how vector interactions, finite spectral widths, and beyond-mean-field non-Gaussian correlations modify the reduced correlator and its entropy. Extensions to PNJL-type or other confining frameworks would also allow one to study the interplay between deconfinement and reduced chirality entropy in a more realistic setting. On the nonperturbative side, future lattice-QCD studies based on fermionic two-point functions and equal-time correlation matrices may help assess the extent to which chirality-projected reduced entropies can serve as robust information-theoretic probes of chiral dynamics in strongly interacting matter.

\section*{Acknowledgment}
This work was supported by the National Research Foundation of Korea (NRF), funded by the Korean government (MSIT) (RS-2025-16065906).

\section*{Appendix}
\subsection{Validity range of the equilibrium assumption}
The equilibrium relations used in Eqs.~(\ref{eq:GGGG}) and (\ref{eq:eq_relation_main}),
\begin{align}
G^{<}(p) = i\,n_F(p^0)A(p), \quad
G^{>}(p) = -i\,[1-n_F(p^0)]A(p),
\label{eq:eq_relation}
\end{align}
are valid when the system is in local thermal equilibrium and its relaxation time $\tau_{\mathrm{rel}}$ is sufficiently longer than the microscopic interaction time. In this regime, quasiparticle excitations are well defined, and the spectral function is sharply peaked near the mass shell,
\begin{equation}
A(p) \simeq 2\pi\,\mathrm{sgn}(p^0)\,\delta(p^2-M_q^2),
\label{eq:spectral_narrow}
\end{equation}
so that finite-width effects can be neglected. This narrow-width quasiparticle limit justifies the use of the on-shell expressions in Eqs.~(\ref{eq:C_equal_time_main})--(\ref{eq:Ck_final}).

If a finite-width spectral function is retained, the equal-time correlator is still defined by
\begin{equation}
C(\bm{k}) = i\int \frac{dk^0}{2\pi}\,G^{<}(k),
\label{eq:appendix_C}
\end{equation}
and the chirality-reduced correlator by
\begin{equation}
C_L(\bm{k}) = P_L C(\bm{k}) P_L
= i\int \frac{dk^0}{2\pi}\,P_L G^{<}(k) P_L.
\label{eq:appendix_CL}
\end{equation}
Accordingly, $S_\chi$ measures the mixedness of the reduced chiral sector including possible broadening effects, and reduces to the on-shell expression in Eq.~(\ref{eq:entropy_main_revised}) in the narrow-width limit. In this sense, finite spectral widths provide an additional source of reduced-sector mixedness associated with thermal smearing and quasiparticle damping.

\subsection{Justification for using the equal-time reduced correlator}
In equilibrium, the lesser Green's function $G^{<}(k)$ describes the occupied part of the single-particle spectrum, whereas the entropy of a Gaussian fermionic subsystem is determined by the corresponding equal-time correlation matrix. For this reason, we define
\begin{equation}
C(\bm{k}) = i\int \frac{dk^0}{2\pi}G^{<}(k),
\quad 0 \le C(\bm{k}) \le 1.
\label{eq:appB_Ck}
\end{equation}
For a Gaussian (mean-field) fermionic state, this correlator completely specifies the state at the two-point level.

Projecting onto the left-handed subspace with $P_L=(1-\gamma^5)/2$, we obtain
\begin{equation}
C_L(\bm{k}) = P_L C(\bm{k}) P_L
= i\int \frac{dk^0}{2\pi}\,P_L G^{<}(k) P_L.
\label{eq:appB_CL}
\end{equation}
This $C_L(\bm{k})$ is the reduced equal-time correlation matrix for the left-handed sector. Its eigenvalues satisfy
\begin{equation}
0 \le \lambda_\pm(\bm{k}) \le 1,
\end{equation}
so they admit the standard fermionic probability interpretation and determine the entropy of the reduced Gaussian subsystem. The corresponding von Neumann entropy is
\begin{equation}
S_\chi=-\mathrm{Tr}\left[C_L\ln C_L+(1-C_L)\ln(1-C_L)\right].
\label{eq:appB_entropy}
\end{equation}
Thus, within the mean-field or quasiparticle approximation, $C_L$ provides the correct object for the entropy construction.


\end{document}